\documentclass[superscriptaddress,twocolumn,preprintnumbers,amsmath,amssymb]{revtex4}

\usepackage{graphicx}
\usepackage{dcolumn}
\usepackage{bm}
\usepackage{amsmath}
\usepackage{epsfig}
\usepackage{rotating}
\usepackage{enumerate}
\usepackage{xcolor}

\newcommand{\degree}{\ensuremath{^\circ}}

\begin{document}

\title{Phonon transport across twin boundaries and twin superlattices}

\author{Kim L\'opez-G\"uell}
\affiliation{Institut de Ci\`encia de Materials de Barcelona, ICMAB--CSIC,
             Campus UAB, 08193 Bellaterra, Spain}

\author{Nicolas Forrer}
\affiliation{Departement Physik, Universit\"at Basel,
             Klingelbergstrasse 82, 4056 Basel, Switzerland}

\author{Xavier Cartoix\`a}
\affiliation{Departament d'Enginyeria Electr\`onica,
             Universitat Aut\`onoma de Barcelona,
             08193 Bellaterra, Barcelona, Spain}

\author{Ilaria Zardo}
\affiliation{Departement Physik, Universit\"at Basel,
             Klingelbergstrasse 82, 4056 Basel, Switzerland}

\author{Riccardo Rurali}
\affiliation{Institut de Ci\`encia de Materials de Barcelona, ICMAB--CSIC,
             Campus UAB, 08193 Bellaterra, Spain}
\email{rrurali@icmab.es}

\date{\today}

\begin{abstract}
Crystal phase engineering
gives access to new types of superlattices where, rather than different materials,
different crystal phases of the same material are juxtaposed. Here, by means
of atomistic nonequilibrium molecular dynamics calculations, we study to what
extent these periodic systems can be used to alter phonon transport, similarly
to what has been predicted and observed in conventional superlattices based
on heterointerfaces. We focus on twin superlattices in GaAs and InAs and highlight
the existence of two different transport regimes: in one each interface behaves
like an independent scatterer; in the other, a segment with a sufficiently large
number of closely-spaced interfaces, is seen by propagating phonons as a metamaterial
with its own thermal properties.
\end{abstract}

\maketitle

\section{Introduction}
\label{sec:intro}

The design of materials with tailor-made thermal properties is very attractive 
for several applications, ranging from efficient thermoelectrics~\cite{ZebarjadiEES12,
BenentiPR17} to thermal management~\cite{MooreMT14}. A way to engineer the 
phonon spectrum of a material, and thus to tune its thermal conductivity, 
is by creating superlattices, where wave interference creates forbidden 
energy bandgaps for phonons~\cite{MaldovanNatMat15}.  
An additional interest in superlattices is that they allow, in principle, 
to observe the crossover from a particle- to a wave-like phonon transport 
regime, a topic of both fundamental and applied importance. When phonons 
travel across far apart interfaces, they are better
described as particles that suffer multiple independent diffusive scattering events,
each one characterized by the thermal boundary resistance of that 
interface~\cite{SwartzRMP89,RuraliPCCP16}. When the number of 
interfaces or their density increases, interference effects can build up 
and heat transport is better understood by taking into account the wave-nature
of phonons. In the first situation the thermal conductance is tuned by 
controlling the number of interfaces; once the coherent regime kicks-in,
on the other hand, the main control knob becomes the periodicity of the 
superlattice, which, in turn, determines the details of the phonon 
dispersion of the metamaterial, including the position and width of 
the phonon bandgap.
This transition typically occurs by making the interface spacing  
of the same order of the phonon mean free path~\cite{SimkinPRL00},
a goal that can be achieved either increasing the interface density
(i.e. reducing the superlattice period) or decreasing the 
temperature~\cite{YangPRB03,GargPRB13}.
Experimental indications of these effects have been reported in
GaAs/AlAs~\cite{LuckyanovaScience12} and 
perovskite oxides superlattices~\cite{RavichandranNatMat13}.
In this scenario, the quality 
of each individual interface is almost as important as their periodic 
arrangement~\cite{KohAdvFunctMat09}. Indeed, atomic-scale 
corrugations~\cite{RufPRB94,WenPRL09,TianPRB14,MeiJAP15} 
and interfacial chemical mixing~\cite{LandryPRB09b,HubermanPRB13} 
have been shown to largely suppress the coherence of phonon 
transport. Therefore, the design of unconventional periodic
structures, beyond the usual heterostructured superlattices, 
is attracting a considerable attention~\cite{FrielingAPL14}.

\begin{figure}[t]
\includegraphics[width=1.0\linewidth]{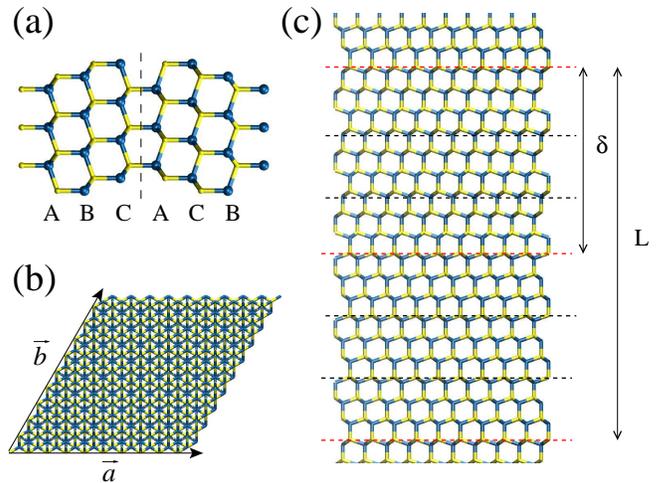}
\caption{(a)~Zoomed view of a twin defect, showing the change from
         ABC to ACB stacking. (b)~Cross-section view of the computational
         cell, where $a=b=39.97$~\AA for GaAs and 42.85~\AA\ for InAs. 
         (c)~Segment of a TSL containing
         one full period, $L = 2\delta$, where $\delta$ is the separation
         between adjacent twins. The red dashed lines indicate the position
         of the twins, while the black dashed line show the unit cell
         of the zincblende, when the [111] crystal axis is taken to be
         parallel to the cartesian z-axis. Blue spheres represent Ga or In
         atoms and yellow spheres represent As atoms.
         .}
\label{fig:struct}
\end{figure}

The increased control in the growth of semiconducting nanowires 
(NWs)~\cite{WangMatSciEng08,CaroffIEEESTQE11} has brought attention 
to a novel type of superlattice, where, rather than chemically different 
materials, different crystal phases of the same material --typically 
zincblende and wurtzite segments of arsenides and phosphides-- alternate 
with a well-defined periodicity. These crystal phase 
superlattices~\cite{XiongNL06,CaroffNatureNano09,DickNL10} are made of 
{\it homojunctions} that have (i)~a minimal lattice mismatch and 
(ii)~no chemical intermixing, and thus are virtually atomically flat. 
An even subtler kind of periodic superstructure that can be obtained 
in NWs is the so-called twin superlattice (TSL)~\cite{AlgraNature08,
BurgessACSNano13}. Here both the material and the crystal phase are 
the same throughout the entire NW length and the motif of the superlattice 
is determined by a periodic rotation of the crystal lattice, which
induces the formation of a stacking planar defect that we 
refer to as twin boundary. TSLs of practical interest are made of 
zincblende materials that feature a 60\degree\ rotations of the crystal 
lattice, so that the ABC stacking along the [111] crystal axis switches 
to ACB after each twin boundary (see the sketch in Figure~\ref{fig:struct}a);
see Ref.~\citenum{WoodNanoscale12} for other types of
twin boundaries.

Twin boundaries are peculiar interfaces under many respects
and defy the most common phenomenological
approaches to the calculation of the thermal boundary resistance
(TBR), i.e. the thermal resistance of an interface. The popular
acoustic mismatch model (AMM) computes the TBR in terms of the
mismatch of elastic properties of the constituent materials forming 
an interface. The rationale behind it is that a phonon impinging 
on the interface from one side is efficiently transmitted
only if a suitable vibrational state, in terms of energy and 
momentum, exists on the other side. In a twin boundary, however,
the materials on the two sides of the interface are identical
and the AMM predicts no TBR, at odds with experimental
results that have convincingly shown that they do have an effect on
phonon dispersion~\cite{DeLucaNL19} and with theoretical atomistic 
simulations of phonon transport~\cite{PorterPRB16,CarreteNanoscale19}.
The diffuse mismatch model (DMM), on the other hand, returns the 
same finite value of the TBR that it would erroneously attribute to a 
homogeneous system with no interface~\cite{dmm}.
Previous theoretical calculations, combining molecular dynamics and 
Green's functions, showed that the TBR of twin boundaries 
in GaP and InP is determined by the rotation of the phonon 
polarization vectors and by local structural distortion at the 
interface~\cite{CarreteNanoscale19}. This explains why simplified
models based on the mismatch of the elastic properties cannot capture
phonon scattering at twin boundaries and call for modeling
approaches that explicitly account for the atomic structure 
of the interface.

\section{Computational Methods}
\label{sec:methods}

We perform non-equilibrium molecular dynamics (NEMD) simulations
with the LAMMPS code~\cite{PlimptonJCP95} and a Tersoff-type
interatomic potential~\cite{TersoffPRB89} parameterized by 
Nordlund and coworkers~\cite{NordlundCMS00}. We consider 
GaAs and InAs computational cells with the transport direction
parallel to the cubic [111] crystal axis, which we take to
be the $z$ coordinate direction. This choice is dictated
by the fact that this is growth direction of nanowires
along which twin boundaries can be formed during growth.
The ends of the
computational cells are connected to Nose-Hoover thermostats 
at temperatures $T_H$ and $T_C$, while the rest of the system
evolves according to the microcanonical ensemble.
We start from the 6-atom unit cell of 
zincblende crystals, which has the [111] crystallographic
direction parallel to the $z$-axis and construct
$10 \times 10 \times M$ supercells. We take $M=90$ for the
study of isolated interfaces and $M=180$ for systems
featuring multiple interfaces, i.e. a superlattice segment. 
We apply periodic boundary conditions along $x$- and $y$-axis.
The TBR is calculated as $\Delta T/J$, where $\Delta T$
is the temperature jump at the interface and $J$ is the 
heat flux~\cite{SwartzRMP89} (see Refs.~\cite{RuraliPCCP16,
DettoriAiPX16} for a more general discussion).

After the simulation starts, a thermal gradient rapidly builds 
up, but we nevertheless disregard the first $3 \cdot 10^6$
steps to allow a proper equilibration of the system. In
all cases this time interval proved to be sufficient to
reach the nonequilibrium steady-state. Indeed, after this
equilibration interval, not 
only the time evolution of the local temperature along
$z$ is roughly constant, but also the rate of energy injected 
and extracted by the hot and cold thermostats are the same, 
within numerical fluctuations. After the steady-state
is reached, we average the heat flux and the temperature
over runs that go from 7.5 to $30 \cdot 10^6$ steps.
We apply a temperature difference $T_H-T_C$ equal to 100~K,
varying $T_C$ and $T_H$ in order to obtain a different
average temperature, $T_M$. Notice that below the Debye
temperature the use of classical dynamics, where quantum effects
are neglected, should be handled with care. Yet, attempts at
correcting for quantum features in low temperature molecular 
dynamics provided inconclusive and contrasting results~\cite{BedoyaPRB14}.
We address the effect of this limitation in detail below.

NEMD notoriously suffers from finite-size effects and the usual
procedure to estimate the thermal conductivity of a material
requires running simulations in increasingly large cells.
In this work, however, our goal is either computing 
TBRs, which is much less sensitive to cell 
sizes~\cite{RuraliPRB14}, or studying the dependence of the
thermal resistance in multi-interface systems, where we
compare results obtained in cells of the same size.

\begin{figure}[t]
\includegraphics[width=1.0\linewidth]{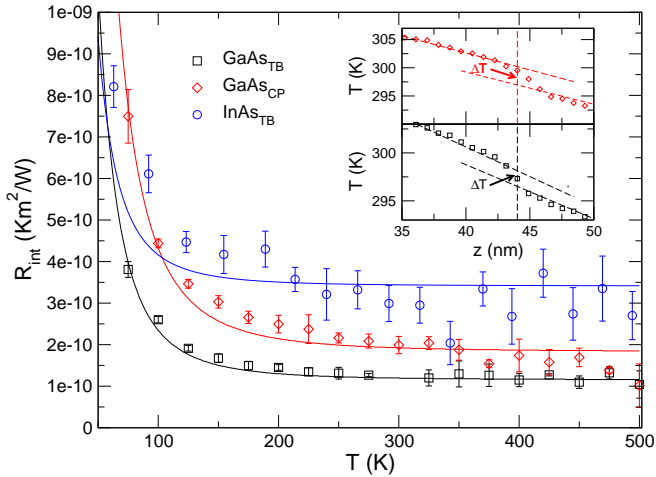}
\caption{TBR as function of temperature of an single twin
         boundary in GaAs and InAs and of a crystal phase
         interface in GaAs. The continuous lines are fits 
         of the computed data to a $T^{-3}$ dependence of the
         TBR. The inset shows the temperature profile,
         $T(z)$ around a crystal phase interface (top) and a twin 
         boundary in GaAs (bottom). The temperature jump, 
         $\Delta T$, that is the signature of the TBR and that is
         used to compute it, is indicated.
         }
\label{fig:TBR}
\end{figure}

\section{Results and discussion}
\label{sec:results}

\subsection{Isolated twin boundary}
\label{sub:1twin}

As a first step of our study we have calculated the TBR
as a function of the interface temperature of a twin boundary
in GaAs and InAs, as displayed in Figure~\ref{fig:TBR}. To this 
end, after reaching the nonequilibrium steady-state, we estimate 
the heat flux from the energy per unit time injected/extracted by 
the hot/cold thermostats and the thermal gradient from the averaged 
temperature profile, $T(z)$; representative examples are provided in the insets.
The results obtained are similar to previous reports of GaP
and InP, showing that the physical effects are general and do
not qualitatively depend on the material or on the classical
potential used to describe interatomic interactions (Vashista
potential for phosphides~\cite{Ribeiro-SilvaJPCM11,BranicioJPCM09}, 
Nordlund for arsenides~\cite{NordlundCMS00}). In the
case of GaAs, we add for comparison a crystal phase interface
between segments of zincblende and wurtzite crystals (red diamonds 
in Figure~\ref{fig:TBR}), which is also a class of important 
interfaces in NW physics~\cite{CaroffNatureNano09,DickNL10}.
In agreement with the results obtained with GaP and InP, we
find a TBR of the order of $2 \times 10^{-10}$~K~m$^2$~W$^{-1}$, 
smaller than conventional heterointerfaces, but slightly larger 
that the one obtained for the corresponding twin boundary (black 
squares in Figure~\ref{fig:TBR}). The continuous lines
are fits of the calculated data to $T^{-3}$, the temperature 
dependence predicted by the same phenomenological models that
fail to account for the TBR~\cite{SwartzRMP89} of twin boundaries. While
this temperature dependence seems to be reasonable for 
both the twin boundary and the crystal phase interface 
in GaAs, in the case of InAs we obtain a somewhat more abrupt
saturation to the high temperature value, preventing a
satisfactory fit to a $T^{-3}$ decay, particularly at low temperatures.
A possible reason is the fact that the Debye temperature of InAs is 
80~K lower than that of GaAs and thus low temperature results
obtained within classical molecular dynamics are comparatively 
less accurate in the former case.

\subsection{Increasing the number of twins}
\label{sub:ntwins}

Now that we have established that individual twin boundaries
in arsenides behave similarly to conventional heterointerfaces,
though with a smaller associated TBR, we move to the study of
multiple interfaces. Our goal is assessing to what extent
TSLs, periodic superstructures made of an ordered sequence of twins, 
behave like conventional superlattices in altering phonon 
transport. In a first set of computational experiments, we 
have considered an increasing number of twin boundaries, 
$1 \leq N \leq 20$, located in the central part of the computational 
cell. The separation between neighboring twins, $\delta$, is fixed, 
so that the twinned region has a thickness equal to $N \delta$. We 
take $\delta=29.4$~\AA\ for GaAs and 31.5~\AA\ for InAs,
i.e. three unit cells along the [111] crystal axis.
Notice that the size of the computational cell along the transport
direction has been doubled with respect to the one used for the
results of Figure~\ref{fig:TBR}. In this way we guarantee that the
twinned region is sufficiently separated from the thermostats, also 
for the largest values of $N$.

\begin{figure}[t]
\includegraphics[width=1.0\linewidth]{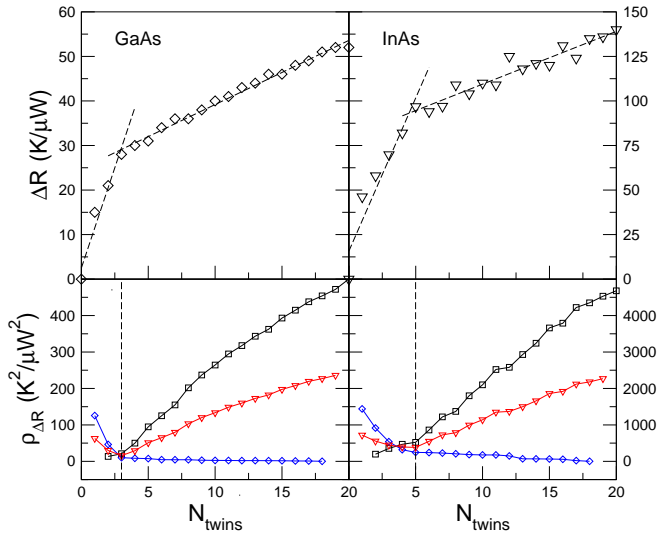}
\caption{(Top) Thermal resistance as a function of the number 
         of twins in GaAs and InAs, at low temperature. $\Delta R$ 
         is defined as $R_N - R_0$, where $R_N$ and $R_0$ are the 
         thermal resistances of a system with $N$ and zero twin 
         boundaries.
         (Bottom) Mean of the residual squares of two linear fits 
         of $\Delta R [0,N]$ and $\Delta R [N,20]$ as a function 
         of $N$ (red triangles); the individual residual squares of 
         each fit is also shown (black squares and blue diamonds).  
         The mean has a minimum at $N=3$ and $N=5$ for GaAs and InAs,
         respectively.
         The corresponding linear fits are shown in the upper panels
         with a dashed line.
         }
\label{fig:ntwinslow}
\end{figure}

\begin{figure}[t]
\includegraphics[width=1.0\linewidth]{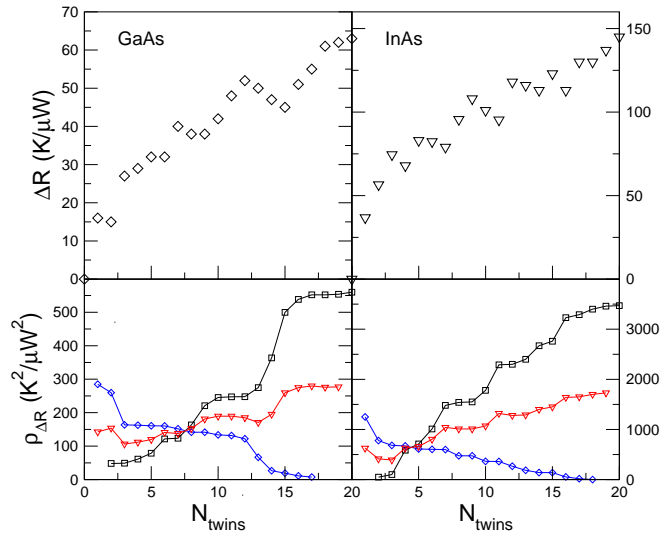}
\caption{(Top) Thermal resistance as a function of the number
         of twins in GaAs and InAs, at room temperature. $\Delta R$
         is defined as $R_N - R_0$, where $R_N$ and $R_0$ are the
         thermal resistances of a system with $N$ and zero twin
         boundaries.
         (Bottom) Mean of the residual squares of two linear fits
         of $\Delta R [0,N]$ and $\Delta R [N,20]$ as a function
         of $N$ (red triangles); the individual residual squares of
         each fit is also shown (black squares and blue diamonds).
         No clear minima emerge.
         }
\label{fig:ntwinshigh}
\end{figure}

We computed $\Delta R = R_N - R_0$, where $R_N$ and
$R_0$ are the thermal resistances of a system with $N$ and zero twin boundaries.
The thermal resistances are computed as $I/ \Delta T$, where $I$ is the heat 
current and $\Delta T$ is evaluated between the values of the $z$ coordinate 
axis $z_i$ and $z_f$, taken to be sufficiently
far from the thermal reservoirs to avoid the usual non linearities of the
temperature. We considered two different values of the
average temperature $T_M = (T_H + T_C)/2$:
a low temperature case, with $T_M = 100$~K
and a higher temperature $T_M=300$ and 250~K, for GaAs and InAs, respectively,
which is of the order of the Debye temperature of each material
($T_D=360$~K for GaAs and $T_D=280$~K for InAs).

The results for GaAs and InAs at $T_M = 100$~K are shown in Figure~\ref{fig:ntwinslow}.
If each twin boundary acted as an independent scatterer for phonons, $\Delta R$
should increase linearly and its slope would be related to the TBR of 
an individual twin. Indeed, this is what we observe at the beginning 
for the low temperature simulations of both materials (see 
Figure~\ref{fig:ntwinslow}). However, as the 
number of twins increases the behavior changes and a different transport regime
can be identified. Once more than 3-5 twins are stacked next to each other,
a collective effect builds up and the slope of $\Delta R$ changes. 
Notice that a hint of the coherent behaviour that develops in the 
many-twin limit was already present in the low-twin transport regime. 
Indeed, as better seen in the low temperature results of Figure~\ref{fig:ntwinslow}, 
the value of $\Delta R$ for two twins is less than twice the value 
it takes of the individual twin, an indication of the interaction 
between neighbouring twins for the chosen value of $\delta$.
In order to establish in a quantitative way for which value of $N$ the slope changes, we
have carried out two linear fits of the computed $\Delta R$ in the
intervals $[0,N]$ and $[N,20]$, varying $N$, and plotted the mean residual
of squares. We found that the best fits are obtained for $N=3$ 
and $N=5$ for GaAs and InAs, respectively. We also checked if the
fit can be improved significantly by assuming more than two linear
regressions, but this was not the case.

In the higher temperature case, where $T_M=250$ or 300~K,
such a distinction between two 
transport regimes is more difficult to make. While by visual 
inspection it seems that for both materials the slope changes
at $N \sim 3$, no clear minima of the mean of the residual squares 
emerge (see Figure~\ref{fig:ntwinshigh}).

We recall here that at low temperatures, the results obtained
from classical molecular dynamics cannot be taken quantitatively.
Assuming that atoms move according to Newton's laws 
implies that phonon population follows Maxwell-Boltzmann,
rather than Bose-Einstein statistics and this is a good
approximation only at sufficiently high temperatures. Simplified schemes
to correct for these effects include a temperature 
renormalization~\cite{HuNL09,HuAIPCP09,SaizBretinCarbon18,SoleimaniCMS18}
(the results obtained at a nominal temperature $T_{MD}$ are actually valid 
at a different temperature, $T_Q$)
or using quantum, rather than classical specific heat, when
it is required for the calculation of the thermal conductivity~\cite{MelisEPJB14,MelisPRL14}. 
However, even within the more general discussion of Berens 
{\it et al.}~\cite{BerensJCP83}, the result of properly accounting 
for quantum effects is a temperature dependent correction of the
thermodynamical variables calculated that can, for instance,
be responsible for the low temperature behavior of the individual 
InAs twin boundary discussed above. The
results of Figure~\ref{fig:ntwinslow}, however, have all been obtained
at the same temperature and thus any small correction to the 
computed values would affect in a similar way all data points. Clearly,
we are not concerned here with the specific values of $\Delta R$,
but rather on highlighting two different transport regimes.
In this sense, our results are reliable and
provide a valuable insight of the underlying physics.

\subsection{Temperature profiles}
\label{sub:tz}


\begin{figure}[t]
\includegraphics[width=1.0\linewidth]{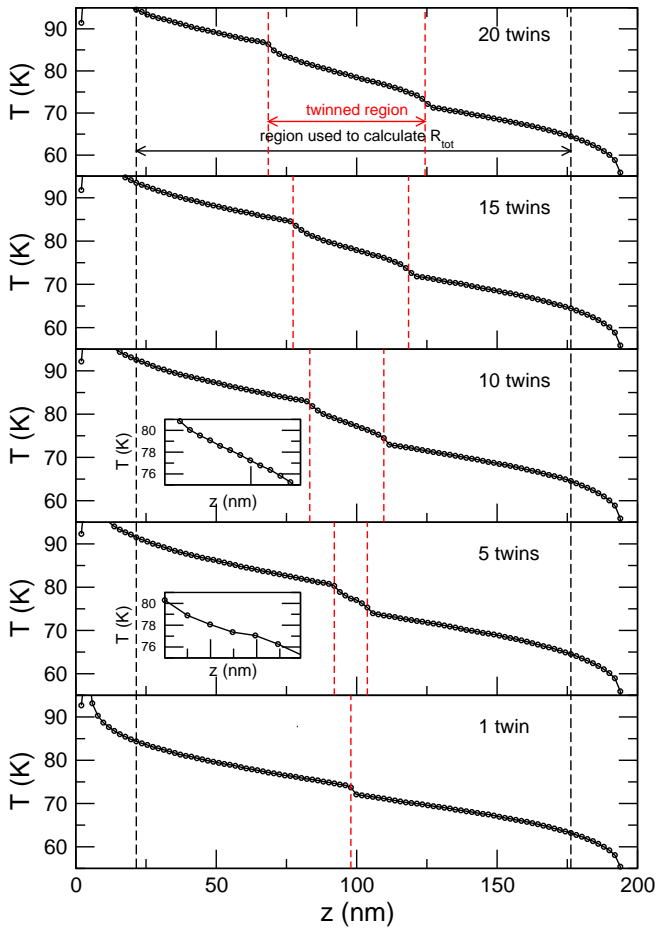}
\caption{Temperature profile, $T(z)$ for a GaAs system with
         $N=1$, 5, 10, 15, and 20 twins with a constant intertwin
         separation, $\delta$ and with $T_M = 100$~K. The red dashed lines indicate
         the position of the first and last twin boundary (i.e., 
         the thickness of the twinned region). The coordinates
         $z_i$ and $z_f$, used to evaluate $\Delta T$ to yield
         $\Delta R$ in Figure~\ref{fig:ntwinslow} are marked
         by the red dashed lines. Inset display a zoomed view
         of $T(z)$ of the twinned region. 
         }
\label{fig:ntemp}
\end{figure}

The spatial dependence of the temperature along the transport
direction is essential in NEMD not only to estimate the TBR, but 
also to compute the thermal conductivity of a homogeneous system.
Indeed, even when the temperature of the reservoirs is fixed, 
it is safer to rely on the thermal gradient of a central region 
of the simulation cell, rather than on the nominal thermal bias, 
$T_H - T_C$. Additionally, $T(z)$ conveys an important physical
insight in presence of interfaces. In 
Figure~\ref{fig:ntemp} we plot $T(z)$ for a GaAs system with 1, 5, 10,
15, and 20 twin boundaries at $T_M=100$~K. The case with one single twin
resembles the temperature profile shown in the inset of
Figure~\ref{fig:TBR}, although here results have been obtained in
a larger computational cell. The temperature discontinuity 
is small, but clearly observable and it occurs exactly where 
the twin boundary is located. Next, we consider the case of 
five twin boundaries. Although it is not possible to observe 
the signature of the five interfaces, it is clear that within 
the twinned region $T(z)$ still exhibits a certain structure
(see the zoomed view in the inset). In the rest of the cases
displayed here ($N > 5$), on the other hand, the twinned regions behaves at all effects 
as a {\it metamaterial} with its own thermal resistance. Phonons 
experience a TBR between pure GaAs segments and a central TSL 
segment. Remarkably, the TBR to enter/exit the TSL region are,
though small, clearly visible and similar in all the cases.
Conversely, within the twinned region $T(z)$ is linear, as
expected in a homogeneous material (see the inset for $N=10$).
See the Supporting Information for the results of GaAs 
at $T_M=300$~K and for InAs at $T_M=100$ and 250~K.

The analysis of Figure~\ref{fig:ntemp} helps rationalize the
dependence of $\Delta R$ presented above. 
Once the twinned
region is seen by the propagating phonons as a 
segment of TSL with its own resistivity, $\rho_{TSL}$,
the total resistance is simply
\begin{equation}
R = \frac{1}{A} \int_{z_i}^{z_i^t} \rho_{GaAs}[T(z)] dz + 
    \frac{1}{A} \int_{z_i^t}^{z_f^t} \rho_{TSL}[T(z)] dz+ 
    \frac{1}{A} \int_{z_f^t}^{z_f} \rho_{GaAs}[T(z)] dz
\end{equation}
where $L_z=z_f-z_i$ is the total length probed, $A$ is the cross-section, 
$\rho_{GaAs}$ is the temperature dependent thermal resistivity of GaAs; the initial and final 
coordinates of the TSL segment are $z_i^t = L_z/2-N\delta/2$ and $z_f^t 
= L_z/2+N\delta/2$. If, for simplicity, we drop the temperature dependence 
of $\rho_{TSL}$ and $\rho_{GaAs}$, the resistance simply reads 
$R = N \delta \rho_{TSL}/A + (L_z - N \delta ) \rho_{GaAs} / A$.
Therefore, if we look back at Figure~\ref{fig:ntwinslow}, we can
distinguish two transport regimes: at first $\Delta R$ increases 
because the number of twin interfaces increases; next it increases 
because the length of the more resistive TSL segment increases.
In the first case phonons see the twin boundaries as individual
scatterers and the increase of $\Delta R$ is dictated by the
TBR of each twin; in the second case the twinned region is seen 
as a TSL with its own conductivity, which depends on the superlattice
design parameters, e.g. the period.

These considerations suggest that a different balance between the
TBR of the individual interface and the resistivity of the ideal
superlattice, i.e. with a very large $N$, could result in a
different dependence of $\Delta R$ on the number of twins. In
particular, if the TSL was considerably less resistive than the 
pure untwinned systems, after the collective
interface behavior shows up should first 
hit a maximum and then decrease.

\subsection{Increasing the density of twins}
\label{sub:dtwins}

\begin{figure}[t]
\includegraphics[width=1.0\linewidth]{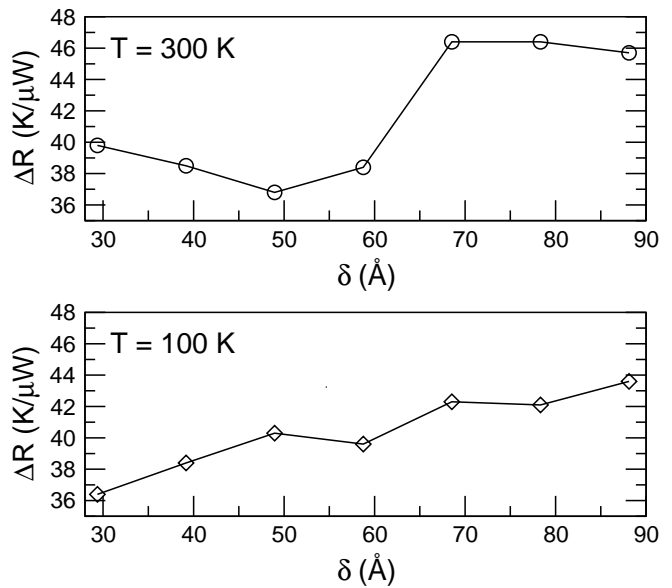}
\caption{Thermal resistance as a function of the intertwin separation,
         $\delta$, for a GaAs system with 6 twin boundaries and
         for (top) $T_M = 300$~K and (bottom) 100~K.
         }
\label{fig:dtwins}
\end{figure}

Finally, we report the results obtained in a different kind of computational
experiments, in order to further corroborate our conclusions.
Now, rather than varying the number of twins and fixing
their separation, we do the opposite: we consider a fixed number 
of six twin boundaries and vary gradually the separation between
neighboring interfaces. If the transport regime was fully incoherent
and each interface scattered phonons independently from the others,
the total thermal resistance should be constant, as all the systems 
contain the same number of twin boundaries. Our results for GaAs 
are displayed in Figure~\ref{fig:dtwins} and capture a rather different 
situation. While for large separations ($\delta > 60$~nm) the 
additional resistance introduced by the twin boundaries is almost
constant, when they are brought together it decreases, a clear 
indication that $\Delta R$ does not simply result from the
sum of independent scattering events. It is interesting that
here the trend is quite clear also for room temperature case,
while such an observation was not fully conclusive when we
varied the number of twin defects (see Figure~\ref{fig:ntwinshigh}).

\section{Conclusions}
\label{sec:concl}

In summary, we have carried out computational experiments based
on nonequilibrium molecular dynamics and provided clear fingerprints
of the fact that twin superlattices behave similarly to their
conventional counterparts, where segments of different materials 
repeat periodically. Namely we have shown that (i)~isolated twin
defects have a small, but finite thermal boundary resistance;
(ii)~when a sufficiently high number of twin defects are stacked
close to each other, phonons see the twinned region as an homogeneous
(meta)material; (iii)~the way a given number of twin defects scatter
phonons depends on their density, indicating the so-called particle
to wave crossover.    
These observations corroborate recent experimental reports~\cite{DeLucaNL19}
and indicate that crystal-phase engineering can become an effective 
way to design materials with desired phononic properties and
to manipulate phonons similar to conventional superlattices, but with
additional advantages, such as defect-free and atomically abrupt
interfaces.

\begin{acknowledgments}
KLG and RR acknowledge financial support by the Agencia
Estatal de Investigaci\'on under grant FEDER-MAT2017-90024-P, and the
Severo Ochoa Centres of Excellence Program under grant CEX2019-000917-S,
and by the Generalitat de Catalunya under grant no. and 2017 SGR 1506.
IZ acknowledges financial support by the European Research Council 
(756365) and the Swiss National Science
Foundation research grant (project grant no. 184942).
We thank the Centro de Supercomputaci\'on de Galicia (CESGA) for the
use of their computational resources.
\end{acknowledgments}

\bibliography{/Users/rrurali/Research/on-going/bibtex/complete}

\begin{thebibliography}{46}
\bibitem[{dmm()}]{dmm}
\bibinfo{howpublished}{This is a known shortcoming of the DMM. If one considers
  a homogeneous system, without any physical interface, and applies the recipe
  of the DMM across a fictitious boundary, a finite TBR is obtained.}
\end{thebibliography}

\end{document}